\documentclass[final,twocolumn,epjc3]{svjour3} 

\smartqed  % flush right qed marks, e.g. at end of proof

\RequirePackage{graphicx}
\RequirePackage{mathptmx}      % use Times fonts if available on your TeX system
\RequirePackage{flushend}
\RequirePackage[numbers,sort&compress]{natbib}
\RequirePackage[colorlinks,citecolor=blue,urlcolor=blue,linkcolor=blue]{hyperref}
\RequirePackage{ulem}

\usepackage{amssymb}
\usepackage{amsmath}
\usepackage{amsfonts}%
\begin{document}
%  \preprint{\hbox{IDSTU-JINR-2012}}

\title{The light-by-light contribution to the muon (g-2) from lightest pseudoscalar and scalar mesons within nonlocal chiral quark model.}
\author{
        A.~E.~Dorokhov\thanksref{eD,addrJINR,addrMSU}
        \and
        A.~E.~Radzhabov\thanksref{eR,addrIDSTU} %etc.
        \and
        A.~S.~Zhevlakov\thanksref{eZ,addrIDSTU} %etc.
}

\thankstext{eD}{e-mail: dorokhov@theor.jinr.ru}
\thankstext{eR}{e-mail: aradzh@icc.ru}
\thankstext{eZ}{e-mail: zhevlakov1@gmail.com}

\institute{ Bogoliubov Laboratory of Theoretical Physics, JINR, 141980 Dubna, Russia\label{addrJINR}
          \and
          N.N.Bogoliubov Institute of Theoretical Problems of Microworld, M.V.Lomonosov Moscow State University, Moscow 119991, Russia\label{addrMSU}
          \and
          Institute for System Dynamics and Control Theory SB RAS, 664033 Irkutsk, Russia \label{addrIDSTU}
}

\date{Received: date / Accepted: date}

\maketitle

\begin{abstract}
The light-by-light contribution from the lightest neutral pseudoscalar and scalar mesons
to the anomalous magnetic moment of muon is calculated in the framework of the
nonlocal $SU(3)\times SU(3)$ quark model.
The model is based on
chirally symmetric
four-quark interaction of the Nambu--Jona-Lasinio type and
Kobayashi--Maskawa--t`Hooft
$U_A(1)$ breaking
six-quark interaction.
Full kinematic dependence of vertices with off-shell mesons and photons in intermediate states in the
light-by-light scattering amplitude is taken into account. The small positive contributions from the 
scalar mesons
stabilize the total result with respect to change of model parameters and reduces to
$a_{\mu }^{\mathrm{LbL,PS+S}}=(6.25\pm 0.83)\cdot 10^{-10}$.
\end{abstract}

\section{Introduction}

The description of the muon anomalous magnetic moment (AMM) is one of the most challenging problem of the elementary particle physics.
Recent precise results on the muon AMM obtained in the experiment E821 at BNL \cite{Bennett:2006fi} open possibility for
very fine investigation of the contributions from the electromagnetic, weak and strong sectors of the standard model.
At present, the theoretical predictions, based on $e^+e^-$ annihilation and $\tau$ decay inclusive cross sections, underestimate the experimental result by approximately 3$\sigma $ (see, e.g. \cite{Jegerlehner:2009ry,Davier:2010nc,Jegerlehner:2011ti}).

The most problematic part of the theoretical estimates is the contribution of the light-by-light (LbL) scattering thro\-ugh the hadronic vacuum.
The LbL scattering contribution (unlike to the hadronic vacuum polarization contribution) can not be extracted from phenomenological considerations or calculated from first principals.
Different models were used for calculation of the LbL contribution to the muon AMM (see, e.g.,
\cite{deRafael:1993za,Hayakawa:1995ps,Bijnens:1995cc,Hayakawa:1997rq, 
Pivovarov:2001mw,
Knecht:2001qf,Melnikov:2003xd,Nyffeler:2009tw,Cappiello:2010uy,
Bijnens:1995xf,Bartos:2001pg,
Dorokhov:2008pw,Dorokhov:2011zf, 
Fischer:2010iz,Goecke:2010if, 
Hong:2009zw}).

In \cite{Dorokhov:2008pw,Dorokhov:2011zf} the contribution of the diagrams with pseudoscalar meson exchanges to the muon AMM was estimated in
the framework of nonlocal chiral quark model (N$\protect\chi$QM) (for details see, e.g., \cite{Anikin:2000rq,Dorokhov:2003kf,Radzhabov:2003hy,Radzhabov:2003hyb,Radzhabov:2010dd} and Appendices of the present work).
In the present paper the contribution of scalar meson exchanges is considered.

Section 2 contains the description of the nonlocal chiral quark model, the meson dynamics in the pseudoscalar and scalar channels with mixing scheme and interaction with external gauge field (see also Appendices A and B). The calculation of the LbL contribution to the muon AMM from the pseudoscalar and scalar exchanges are detailed in Section 3 (see also Appendices C, D and E). Our conclusions are given in Section 4.

\section{\textbf{N$\protect\chi$QM Lagrangian, $T$-matrix and $\protect\eta-%
\protect\eta^{\prime}$ mixing}}

The Lagrangian of the nonlocal $SU(3)\times SU(3)$ model is
\begin{align}
& \mathcal{L}=\bar{q}(x)(i\hat{\partial}-m_{c})q(x)+\frac{G}{2}%
\Big[J_{S}^{a}(x)J_{S}^{a}(x)+J_{P}^{a}(x)J_{P}^{a}(x)\Big] - \nonumber\\
& -\frac{H}{4}%
T_{abc}\Big[J_{S}^{a}(x)J_{S}^{b}(x)J_{S}^{c}(x)-3J_{S}^{a}(x)J_{P}^{b}(x)J_{P}^{c}(x)\Big],
\label{L}
\end{align}%
where $q\left( x\right) $ is the quark field, $m_{c}$ is the diagonal
matrix of the quark current masses\footnote{%
We consider the isospin limit $m_{c,u}=m_{c,d}\neq m_{c,s}$.}, $G$ and $H$
are the four- and six-quark coupling constants. Last line in the Lagrangian
represents the
Kobayashi--Maskawa--t`Hooft determinant vertex \cite{Kobayashi:1970ji,'tHooft:1976fv} with the
structural constant
\begin{equation}
T_{abc}=\frac{1}{6}\epsilon _{ijk}\epsilon _{mnl}(\lambda _{a})_{im}(\lambda
_{b})_{jn}(\lambda _{c})_{kl},
\end{equation}%
where $\lambda _{a}$ are the Gell-Mann flavor matrices for $a=1,..,8$ and $%
\lambda _{0}=\sqrt{2/3}I$. The nonlocal quark currents are
\begin{equation}
J_{ch}^{a}(x)=\int d^{4}x_{1}d^{4}x_{2}\,f(x_{1})f(x_{2})\,\bar{q}%
(x-x_{1})\,\Gamma _{ch}^{a}q(x+x_{2}),  \label{Jx}
\end{equation}%
where $ch=S,P$ and $\Gamma _{{S}}^{a}=\lambda ^{a}$ for the scalar channel, $\Gamma _{{P}%
}^{a}=i\gamma ^{5}\lambda ^{a}$ for the pseudoscalar channel, and $f(x)$ is a form
factor reflecting the nonlocal properties of the QCD vacuum as it occurs in
the instanton liquid model.

The model can be bosonized using the stationary phase approximation which
leads to the system of gap equations for the dynamical quark masses  $m_{d,i}$ $(i=u,d,s)$
\footnote{Through the paper the capital letters will be used for Euclidean momenta, while small letters for Minkowski momenta.}
\begin{align}
& m_{d,u}+GS_{u}+\frac{H}{2}S_{u}S_{s}=0,  \nonumber\\
&m_{d,s}+GS_{s}+\frac{H}{2}%
S_{u}^{2}=0,   \nonumber\\
& S_{i}=-8N_{c}\int \frac{d_E^{4} K}{(2\pi )^{4}}\frac{%
f^{2}(K^{2})m_{i}(K^{2})}{D_{i}(K^{2})},\label{GapEqs}
\end{align}%
where $m_{i}(K^{2})=m_{c,i}+m_{d,i}f^{2}(K^{2})$, $D_{i}(K^{2})=K^{2}+m_{i}^{2}(K^{2})$,
$f(K^{2})$ is the nonlocal
form factor in the momentum representation.

The vertex functions and the meson masses can be found from the
Bethe-Salpeter equation. For the separable interaction, given by Eqs. (\ref{L}), (\ref{Jx}),
the quark-antiquark scattering $T$-matrix in the pseudoscalar (scalar) channel
becomes
\begin{align}
& \mathbf{T}_{ch}=
\hat{\mathbf{T}}_{ch}(P^{2})\delta ^{4}\left( P_{1}+P_{2}-P_{3}-P_{4}\right) \prod\limits_{i=1}^{4}f(P_{i}^{2}),  \nonumber\\
& \hat{\mathbf{T}}_{ch}(P^{2})=\Gamma _{ch}^{k}\left( \frac{1}{-\mathbf{G}_{ch}%
^{-1}+\mathrm{\Pi}_{ch}(P^{2})}\right) _{kl}\Gamma _{ch}^{l},\label{T}
\end{align}%
where $P_{i}$ are the momenta of external quark lines, $\mathbf{G}_{ch}$ and
${\mathrm{\Pi}_{ch} }(P^{2})$ are the corresponding matrices of the four-quark
coupling constants and the polarization operators of mesons 
($P=P_{1}+P_{2}=P_{3}+P_{4}$). The meson masses $\mathrm{M}_{{M}}$ are determined
from the zeros of determinant, $\mathrm{det}(\mathbf{G}_{ch}^{-1}-{\mathrm{\Pi}}_{ch}(-\mathrm{M}_{{M}}^{2}))=0$.
The actual expressions for the matrices $\mathbf{G}_{ch}$ and $\mathrm{\Pi}_{ch}$ are given in \ref{AppA}.

The $\hat{\mathbf{T}}$-matrix for the system of mesons\footnote{%
Such description of the light scalar mesons as $\bar{q}q$-states is probably
simplified. It seems that it is necessary to include other
structures, e.g., four-quark states (see, e.g., \cite{Achasov:2010fh}).
However, the present model is formulated in the leading order of the $1/N_{c}$ expansion
and our calculations are consistent within given approximation. Moreover, the scalar mesons participate in the processes under consideration only as
intermediate states, being far from mass-shell.} in each neutral channel can be
expressed as
\begin{equation}
\hat{\mathbf{T}}_{ch}(P^{2})=\sum_{M}\frac{\overline{V}_{M}(P^{2})\otimes
V_{M}(P^{2})}{-(P^{2}+\mathrm{M}_{{M}}^{2})},
\label{Tch}
\end{equation}%
where $\mathrm{M}_{M}$ are the meson masses, $V_{M}(P^{2})$ are the vertex functions $%
\left( \overline{V}_{M}(p^{2})=\gamma ^{0}V_{M}^{\dag }(P^{2})\gamma ^{0}\right) $.
The sum in (\ref{Tch}) is over full set of light mesons: $(M={\pi ^{0},\eta ,\eta ^{\prime }}%
)$ in the pseudoscalar channel and $(M={a_{0}(980),\sigma,f_{0}(980) })$
in the scalar one. In general case of three unequal quark masses it is
necessary to solve the $\pi ^{0}-\eta -\eta ^{\prime }$ and $%
a_{0}-\sigma -f_{0}$ systems. However, in the isospin limit considered here they reduce to the $%
\pi ^{0}$ and $\eta -\eta ^{\prime }$
systems and to the $a_{0}$ and $\sigma -f_{0}$ systems. Then, it is convenient to diagonalize the scattering
matrix by orthogonal transformations
\begin{align}
\begin{pmatrix}
\eta \\
\eta ^{\prime }%
\end{pmatrix}%
&=%
\begin{pmatrix}
\cos \theta _{{P}} & -\sin \theta _{{P}} \\
\sin \theta _{{P}} & \cos \theta _{{P}}%
\end{pmatrix}%
\begin{pmatrix}
\eta _{8} \\
\eta _{0}%
\end{pmatrix}%
,
 \nonumber\\
\begin{pmatrix}
\sigma \\
f_{0}(980)%
\end{pmatrix}%
&=%
\begin{pmatrix}
\cos \theta _{{S}} & -\sin \theta _{{S}} \\
\sin \theta _{{S}} & \cos \theta _{{S}}%
\end{pmatrix}%
\begin{pmatrix}
\sigma _{8} \\
\sigma _{0}%
\end{pmatrix}%
.  \label{Mix}
\end{align}%

As a result the mesonic vertex functions are
\begin{align}
V_{a_{0}}\left(  P^{2}\right)   &  =ig_{a_0}(P^{2})\lambda
_{3} ,\nonumber\\
V_{\sigma}\left(  P^{2}\right)   &  =ig_{\sigma}(P^{2})\left(
\lambda_{8}\cos\theta_S(P^{2})-\lambda_{0}\sin\theta_S(P^{2})\right),\\
V_{f_0}\left(  P^{2}\right)   &  =ig_{f_0}%
(P^{2})\left(  \lambda_{8}\sin\theta_S(p^{2})+\lambda_{0}\cos\theta_S
(P^{2})\right) ,\nonumber
\end{align}
where $g_{M}(P^{2})$  and $\theta(P^{2})$ are the meson renormalization constants and mixing angles (\ref{AngelMix}) depending on the meson virtuality.
The renormalization constants are defined through the unrenormalized meson
propagators $D_{M}(P^{2})$ as
\begin{align}
{g_{M}^{2}}(P^{2})={-(P^{2}+\mathrm{M}_{{M}}^{2})}D_{M}(P^{2}).
\end{align}

The meson mixing angles depend strongly on the meson virtuality. Therefore $\theta _{\sigma }=\theta _{{S}}(-\mathrm{M}_{\sigma }^{2})$ and $\theta
_{f_{0}}=\theta _{{S}}(-\mathrm{M}_{f_{0}}^{2})$ are different for the on-shell $%
\sigma $ and $f_0$ mesons. The same situation takes place in the
pseudoscalar sector, where $\theta _{\eta }=\theta _{{P}}(-\mathrm{M}_{\eta }^{2})$, $\theta _{\eta ^{\prime
}}=\theta _{{P}}(-\mathrm{M}_{\eta ^{\prime }}^{2})$.

External fields are introduced by delocalization of the quark fields $q(x)$ by using the Schwinger phase factor $E(x,y)$
\begin{align}
q(y)\to Q(x,y) = E(x,y)q(y),
\end{align}
where
\begin{align}
E(x,y)= \mathcal{P}\mathrm{exp}\left\{i\int
\limits_x^y dz^\mu [{\cal V}^a_\mu(z)+ {\cal A}^a_\mu(z)\gamma_5]T^a\right\},
\label{Schwinger}
\end{align}
and ${\cal V}^a_\mu$ and ${\cal A}^a_\mu$ are the external vector and
axial-vector gauge fields, $T^a \equiv \lambda^a/2$.
The $\mathcal{P}\mathrm{exp}$ is handled with help of prescription for the
derivative of contour integral
\begin{align}
   \frac{\partial}{\partial y^{\mu }}\int\limits_{x}^{y}dz^{\nu }\
   F_{\nu}(z)=F_{\mu }(y),\quad \delta^{(4)}\left( x-y\right)
   \int\limits_{x}^{y}dz^{\nu}\ F_{\nu }(z)=0,
\nonumber
\end{align}
as described in \cite{Terning:1991yt}.
As a result the kinetic part leads to usual local electroweak vertices.
However, the terms with nonlocal quark currents $J_{ch}^{a}(x)$ generate additional vertices, see \ref{AppB}.

For numerical estimates we use the Gaussian nonlocal form factor for Euclidean momenta
$f(K^{2})=\exp (-K^{2}/{2\Lambda ^{2}})$ and the model parameters obtained in
\cite{Scarpettini:2003fj}. The model parameters (the current quark masses $%
m_{c,i}$, the coupling constants $G$ and $H$, and the nonlocality scale $%
\Lambda $) are fixed in \cite{Scarpettini:2003fj} by requirement that the model reproduces correctly the
measured values \cite{Nakamura:2010zzi} of the pion and kaon masses, the pion decay constant $f_{\pi
}$, and the $\eta ^{\prime }$ mass (parameter sets $\mathrm{G}_{I}$, $%
\mathrm{G}_{IV}$) or the $\eta ^{\prime }\rightarrow \gamma \gamma $ decay
constant $g_{\eta ^{\prime }\gamma \gamma }$ (sets $\mathrm{G}_{II}$, $%
\mathrm{G}_{III}$). The sets $\mathrm{G}_{I}$, $\mathrm{G}_{IV}$ vary by
different input for the nonstrange current quark mass, while $\mathrm{G}%
_{II} $, $\mathrm{G}_{III}$ are two solutions of the same fitting procedure.

\section{LbL contribution from resonance exchanges}

\begin{figure}[t]
\begin{tabular*}{\columnwidth}{@{}ccc@{}}
\raisebox{-0.5\height}{\resizebox{0.13\textwidth}{!}{\includegraphics{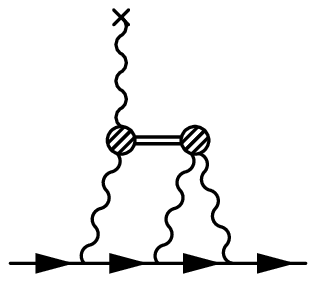}}}&
\raisebox{-0.5\height}{\resizebox{0.13\textwidth}{!}{\includegraphics{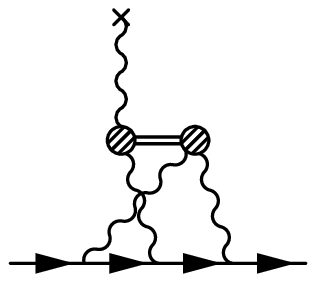}}}&
\raisebox{-0.5\height}{\resizebox{0.13\textwidth}{!}{\includegraphics{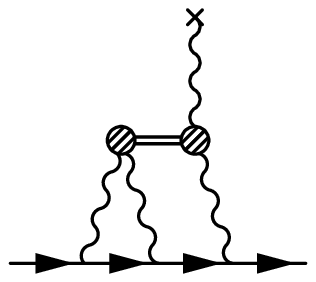}}}\\
(a)&(b)&(c)
\end{tabular*}
\caption{LbL contribution from intermediate meson exchanges.}
\label{fig:LbL}
\end{figure}

The basic element for calculations of the hadronic LbL contribution to the muon AMM is the fourth-rank light quark hadronic vacuum polarization tensor
\begin{align}
&\mathrm{\Pi}_{\mu\nu\lambda\rho}(q_1,q_2,q_3)=\int d^4x_1\int d^4x_2\int d^4x_3 e^{i(q_1x_1+q_2x_2+q_3x_3)}\times\nonumber\\
&\quad\quad\quad\times\left< 0|T(j_\mu(x_1)j_\nu(x_2)j_\lambda(x_3)j_\rho(0))|0\right>,
\end{align}
where $j_\mu(x)$ are light quark electromagnetic currents and $\left|0\right>$ is the QCD vacuum state.

The muon AMM can be extracted by using the projection \cite{Brodsky:1967sr}
\begin{align}
a_\mu^{\mathrm{LbL}}=\frac{1}{48m_\mu}\mathrm{Tr}\left((\hat{p}+m_\mu)[\gamma^\rho,\gamma^\sigma](\hat{p}+m_\mu)\mathrm{\Pi}_{\rho\sigma}(p,p)\right), \nonumber
\end{align}
where
\begin{align}
&\mathrm{\Pi}_{\rho\sigma}(p^\prime,p)=-i e^6 \int \frac{d^4q_1}{(2\pi)^4}\int \frac{d^4q_2}{(2\pi)^4}\frac{1}{q_1^2 q_2^2 (q_1+q_2-k)^2}\times\nonumber\\
&\quad\quad\times \gamma^\mu \frac{\hat{p}^\prime-\hat{q}_1+m_\mu}{(p^\prime-q_1)^2-m_\mu^2}\gamma^\nu \frac{\hat{p}-\hat{q}_1-\hat{q}_2+m_\mu}{(p-q_1-q_2)^2-m_\mu^2} \gamma^\lambda \times \nonumber\\
&\quad\quad \times \frac{\partial}{\partial k^\rho}\mathrm{\Pi}_{\mu\nu\lambda\sigma}(q_1,q_2,k-q_1-q_2), \quad
\end{align}
$m_\mu$ is the muon mass, $k_\mu=(p^\prime-p)_\mu$ and it is necessary to consider the limit $k_\mu \to 0$.

In the case of the resonance exchanges of the light hadrons in the
intermediate pseudoscalar and scalar channel the LbL contribution
to the muon AMM is shown in Fig. \ref{fig:LbL}.
The vertices containing the virtual meson with momentum $p$ and two photons with momenta $q_{1,2}$ and the polarization vectors $\epsilon_{1,2}$
(see \ref{AppC} and Fig. \ref{fig:SggSimp})
can be written as \cite{Bartos:2001pg}%
\begin{align}
&\mathcal{A}\left(  \gamma^{\ast}_{\left(  q_{1},\epsilon_{1}\right)}  \gamma^{\ast}_{\left( q_{2},\epsilon_{2}\right)}  \rightarrow P^{\ast}_{\left(  p\right)} \right)=e^{2}\epsilon_{1}^{\mu}\epsilon_{2}^{\nu}\Delta_P^{\mu\nu}\left( p, q_{1}, q_{2}\right)
\end{align}
with
\begin{align}
\Delta_P^{\mu\nu}\left( p, q_{1}, q_{2}\right)=-i\varepsilon_{\mu\nu\rho\sigma}q_{1}^{\rho}q_{2}^{\sigma} \mathrm{F}_{P}\left(  p^{2};q_{1}^{2},q_{2}^{2}\right) ,
\end{align}
and
\begin{align}
\mathcal{A}\left( \gamma^{\ast}_{\left(  q_{1},\epsilon_{1}\right)}  \gamma^{\ast}_{\left(q_{2},\epsilon_{2}\right) } \rightarrow S^{\ast}_{\left(  p\right)} \right)=e^{2}\epsilon_{1}^{\mu}\epsilon_{2}^{\nu} \Delta_S^{\mu\nu}\left( p, q_{1}, q_{2}\right)
\end{align}
with
\begin{align}
&\Delta_S^{\mu\nu}\left( p, q_{1}, q_{2}\right)= 
\mathrm{A}_{S}\left(p^{2};q_{1}^{2},q_{2}^{2}\right) P_{A}^{\mu \nu }(q_{1},q_{2})+
\label{ApiGG}\\
& \quad \quad \quad \quad \,  +\mathrm{B}_{S}(p^{2};q_{1}^{2},q_{2}^{2})P_{B}^{\mu \nu }(q_{1},q_{2}) 
,\nonumber
\end{align}
where
\begin{align}
&\quad P_{A}^{\mu \nu }(q_{1},q_{2}) =\left( g^{\mu \nu }(q_{1}
q_{2})-q_{1}^{\nu }q_{2}^{\mu }\right) ,\nonumber \\
&\quad P_{B}^{\mu \nu }(q_{1},q_{2}) =\left( q_{1}^{2}q_{2}^{\mu }-(q_{1}
q_{2})q_{1}^{\mu }\right) \left( q_{2}^{2}q_{1}^{\nu }-(q_{1}
q_{2})q_{2}^{\nu }\right) ,\nonumber
\end{align}
and  $p=q_{1}+q_{2}$.
Note, that the scalar form factor $\mathrm{B}_{S}$ is singular in the limit when one photon is real and the virtuality of the second photon equals to the virtuality of the scalar meson $p^{2} \rightarrow q_{1}^{2}$, $q_2^2 \rightarrow 0$.
For convenience we also define an additional function
\begin{align}
\mathrm{B}_{S}^{\prime }(p^{2};q_{1}^{2},q_{2}^{2})=
\mathrm{B}_{S}(p^{2};q_{1}^{2},q_{2}^{2})\left( (q_{1} q_{2})^{2}-q_{1}^{2}q_{2}^{2}\right), \label{BPrime}
\end{align}
which is regular in this limit. 
%}.
In general case, these scalar functions are combinations of the nonstrange and strange components. Details for the mixing of mesons interacting with photons are given in \ref{AppC}. In \ref{AppE} the local limit of the amplitude  $\gamma^{\ast} \gamma^{\ast} \to S^\ast $  is presented.

The
polarization tensor $\mathrm{\Pi}^{\mu\nu\lambda\rho}$ for the exchange of meson with mass $\mathrm{M_M}$ is 
\begin{align}
&\mathrm{\Pi}^{\mu\nu\lambda\rho}(q_1,q_2,q_3)
=\\
&\quad \quad i\frac{\Delta^{\mu\nu    }(q_1+q_2,q_1,q_2        )\Delta^{\lambda\rho}(q_1+q_2,q_3,q_4)}{(q_1+q_2)^2-\mathrm{M_M}^2}+\nonumber\\
&\quad +i\frac{\Delta^{\mu\rho   }(q_2+q_3,q_1,q_4)\Delta^{\nu\lambda }(q_2+q_3,q_2,q_3        )}{(q_2+q_3)^2-\mathrm{M_M}^2}+\nonumber\\
&\quad +i\frac{\Delta^{\mu\lambda}(q_1+q_3,q_1,q_3        )\Delta^{\nu\rho    }(q_1+q_3,q_2,q_4)}{(q_1+q_3)^2-\mathrm{M_M}^2},\nonumber
\end{align}
where $q_i$ are momenta of outgoing photons, $q_4=-(q_1+q_2+q_3)$,
and one should take $\Delta^{\mu\nu}_P$ for pseudoscalar and $\Delta^{\mu\nu}_S$ for scalar mesons, respectively. Details for the pseudoscalar exchange can be found in \cite{{Knecht:2001qf}} (Eqs. (3.1) and (3.3)). The low-energy expansion of the
derivative of the polarization tensor $\mathrm{\Pi}^{\mu\nu\lambda\rho}$ for the scalar meson is given by
\begin{align}
&\frac{\partial}{\partial k^\rho}\mathrm{\Pi}^{\mu\nu\lambda\sigma}(q_1,q_2,k-q_1-q_2)=\nonumber\\
&\quad\, i\frac{\Delta_{S}^{\mu\nu    }(q_1+q_2,q_1,q_2  )}{(q_1+q_2)^2-\mathrm{M_M}^2}\frac{\partial}{\partial k^\rho}\Delta_{S}^{\lambda\sigma}(q_1+q_2,-q_1-q_2,-k)\nonumber\\
&\,+i\frac{\Delta_{S}^{\nu\lambda}(-q_1,q_2,-q_1-q_2)}{ q_1^2     -\mathrm{M_M}^2}\frac{\partial}{\partial k^\rho}\Delta_{S}^{\mu\sigma    }(-q_1,q_1,-k)\\
&\,+i\frac{\Delta_{S}^{\mu\lambda}(-q_2,q_1,-q_1-q_2)}{ q_2^2     -\mathrm{M_M}^2}\frac{\partial}{\partial k^\rho}\Delta_{S}^{\nu\sigma    }(-q_2,q_2,-k)+O(k).\nonumber
\end{align}
and the low-energy expansion for the derivative of $\Delta^{\mu\nu}_S$ is
\begin{align}
&\frac{\partial}{\partial k^\rho} \Delta^{\mu\nu}_S(-q,q,k)=\mathrm{A}_{S}(q^2,q^2,0) \Big(g^{\mu\nu}q^{\rho}-q^{\nu}g^{\mu\rho}\Big) +\nonumber\\
&\quad\,+\mathrm{B}_{S}^\prime(q^2,q^2,0) q^{\nu}\left(\frac{q^{\mu}q^{\rho}}{q^2}-g^{\mu\rho}\right)+O(k).
\end{align}
As a result the numerator of the two-loop integrand for the $a_\mu^{\mathrm{LbL}}$ contains the combination of two form-factors and is a polynomial in momenta.

At next step, the expression for LbL can be averaged \cite{Jegerlehner:2009ry} over directions of
the muon momentum $p$%
\begin{align}
\left\langle ...\right\rangle =\frac{1}{2\pi^{2}}\int d\Omega\left(
\widehat{p}\right)  ...
\end{align}
After averaging 
the expression for the LbL contribution
to the muon AMM from the light scalar meson exchange can be written in the form of integral over Euclidean momenta
\begin{align}
& a_{\mu}^{\mathrm{LbL},\mathrm{S}}=-\frac{2\alpha^{3}}{3\pi^{2}}%
\int\limits_{0}^{\infty}dQ_{1}^{2}\int\limits_{0}^{\infty}dQ_{2}^{2}\int%
\limits_{-1}^{1}dt\sqrt{1-t^{2}}\frac{1}{Q_{3}^{2}}\times   \nonumber\\
&\quad \times\sum_{S=a_0^{0},\sigma,f_0} \biggl[ 2\frac{\mathcal{N}^{S}_{%
\mathbf{1}}}{Q_{2}^{2}+\mathrm{M}_{S}^{2}} +\frac{\mathcal{N}^{S}_{\mathbf{2}}}{%
Q_{3}^{2}+\mathrm{M}_{S}^{2}} \biggr] ,   \\
&\quad \mathcal{N}^{S}_{\mathbf{1}}=
  \sum_{{\mathrm{X=A,B}^\prime}}\sum_{{\mathrm{Y=A,B}}^{}}
  \mathrm{X}_S\left( Q_{2}^{2} ;Q_{2}^{2},0\right)
  \mathrm{Y}_S\left( Q_{2}^{2};Q_{1}^{2},Q_{3}^{2}\right)  \mathrm{Ts}^{\mathrm{XY}}_{\mathbf{1}},\nonumber\\
&\quad \mathcal{N}^{S}_{\mathbf{2}}=
  \sum_{{\mathrm{X=A,B}^\prime}}\sum_{{\mathrm{Y=A,B}}^{}}
  \mathrm{X}_S\left( Q_{3}^{2} ;Q_{3}^{2},0\right)
  \mathrm{Y}_S\left( Q_{3}^{2};Q_{1}^{2},Q_{2}^{2}\right)  \mathrm{Ts}^{\mathrm{XY}}_{\mathbf{2}}, \nonumber%\\
\end{align}
where $X_S$, $Y_S$ are the functions $A_S$, $B_S$, $B^\prime_S$ deined in Eqs. (17), (18), $Q_3=-\left( Q_{1}+Q_{2}\right)$ and capital letters are introduced for Euclidean momenta, i.e. $Q_l^2=-q_l^2$. One should note that $\mathrm{Ts}^{\mathrm{B^\prime A}}_{\mathbf{i}}=\frac{1}{2}\mathrm{Ts}^{\mathrm{AA}}_{\mathbf{i}}$, $\mathrm{Ts}^{\mathrm{B^\prime B}}_{\mathbf{i}}=\frac{1}{2}\mathrm{Ts}^{\mathrm{AB}}_{\mathbf{i}}$.
The functions $ \mathrm{Ts}^{\mathrm{XY}}_{\mathbf{i}}$ are given in \ref{AppD}.

\section{The results of model calculation}

It is instructive to study the $\pi$ and $\sigma$ mesons contribution to the muon AMM 
for the
$SU(2)\times SU(2)$ version of the nonlocal model. The Lagrangian of the model is given by 
\begin{align}
& \mathcal{L}_{SU_2}=\bar{q}(x)(i\hat{\partial}-m_{c})q(x)+\frac{G}{2}%
\Big[J_{S}^{u}(x)J_{S}^{u}(x)+J_{P}^{b}(x)J_{P}^{b}(x)\Big], \nonumber 
\end{align}%
where the corresponding flavor matrices are $\lambda _{u}=(\sqrt{2} \lambda _0 + \lambda _8)/\sqrt{3}=\mathrm{diag}(1,1,0)$ and $b=1,2,3$.
In this case the model has three
parameters: the current quark mass $m_{c,u}$, the dynamical quark mass $m_{d,u}$ and
the nonlocality parameter $\Lambda $. In order to understand the stability of
the model predictions with respect to changes of the model parameters one may
vary one parameter in rather wide physically acceptable interval, while fix
other parameters by using as input the pion mass and the two-photon
decay constant of the neutral pion. Thus, we take the values of the
dynamical quark mass in the typical interval of model values $200$--$350$
MeV and other parameters are fitted by the above physical observables within
the error range given in \cite{Nakamura:2010zzi}.

\begin{figure}[t]
\resizebox{0.45\textwidth}{!}{  \includegraphics{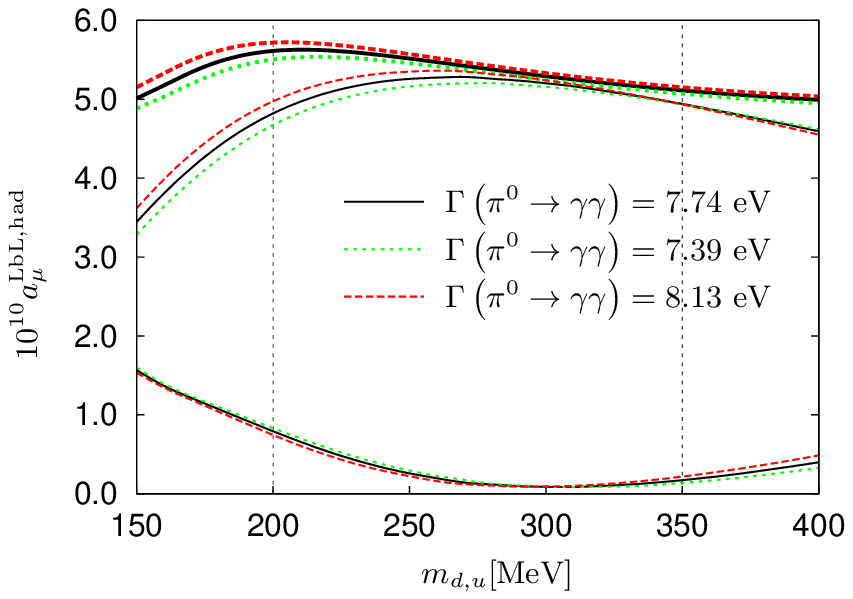}
}
\caption{ LbL contribution to the muon AMM
from the neutral pion and $\protect\sigma $ exchanges as a function of the
dynamical quark mass. Bunch of three lower lines correspond to the $\protect%
\sigma $ contribution, the $\protect\pi ^{0}$ contribution is in the middle,
and the upper lines are the combined contribution. The band along the thick line between
dashed and dotted lines corresponds to the error interval for the pion
two-photon width. Vertical thin dashed lines denote the interval of
dynamical quark masses used for the estimate of the error band for $a_{%
\protect\mu }^{\mathrm{LbL}}$. }
\label{fig:PionLbL}
\end{figure}

The results are shown in Fig. \ref{fig:PionLbL}. We see that the
contribution of the $\sigma $ meson is small, positive and has very small
minimum value around the value of 300 MeV for the dynamical mass. It is interesting to note that
the total result for the pion and $\sigma $ meson contributions is rather
stable to variation of the dynamical mass in the tested interval. Our estimates
for the $\pi ^{0}$ and the sum of $\pi ^{0}$ and $\sigma $ contributions (here and below in
$10^{-10}$) are
\begin{equation}
a_{\mu }^{\mathrm{LbL,\pi ^{0}}}=5.01\pm 0.37,\quad a_{\mu }^{\mathrm{%
LbL,\pi ^{0}+\sigma }}=5.40\pm 0.33.  \label{aSU2}
\end{equation}

\begin{table*}[cth]
\centering
\begin{tabular}{|c|c|c|c|c|c|c|c|c|c|c||c||}
\hline
set & $\pi^{0}$ & $\eta$ & $\eta^{\prime}$ & $\eta+\eta^{\prime}$ & PS & $%
a_0(980)$ & $\sigma$ & $f_0(980)$ & S & $\pi^0+\sigma$ & PS+S \\ \hline
$\mathrm{G}_{I}$ & $5.05$ & $0.55$ & $0.27$ & $0.82$ & $5.87$ & $0.0064$ & $%
0.100$ & $0.0035$ & $0.110$ & $5.15$ & $5.98$ \\
$\mathrm{G}_{II}$ & $5.05$ & $0.59$ & $0.48$ & $1.08$ & $6.13$ & $0.0079$ & $%
0.100$ & $0.0038$ & $0.110$ & $5.15$ & $6.24$ \\
$\mathrm{G}_{III}$ & $5.05$ & $0.53$ & $0.18$ & $0.71$ & $5.76$ & $0.0058$ &
$0.100$ & $0.0034$ & $0.109$ & $5.15$ & $5.87$ \\
$\mathrm{G}_{IV}$ & $5.10$ & $0.49$ & $0.25$ & $0.74$ & $5.84$ & $0.0060$ & $%
0.115$ & $0.0038$ & $0.126$ & $5.25$ & $5.97$ \\ \hline
\end{tabular}%
\caption{The contribution of scalar and pseudoscalar mesons to the muon
AMM $a_{\protect\mu }^{\mathrm{LbL}}$ for different
sets of model parameters \cite{Scarpettini:2003fj}.
All numbers are given in $10^{-10}$.}
\label{table:1}
\end{table*}
In the $SU(3)$ model, for the central values of $\eta $ and $\eta ^{\prime }$ contributions we use
the averages over different parameterizations \cite{Dorokhov:2011zf}. The error
bar for $\eta ^{\prime }$ is taken as a maximal deviation from the central
value. The deviation of the $\eta $ contribution from the central value seems
accidentally small, so we use the factor $60\%$ of the error value for $\eta ^{\prime }$ as an
estimate of the error bar for the $\eta $ contribution. Our estimate for the $\eta $
and $\eta ^{\prime }$ contributions is
\begin{align}
a_{\mu }^{\mathrm{LbL,\eta }}=0.54\pm 0.32,\quad a_{\mu }^{\mathrm{LbL,\eta
^{\prime }}}=0.30\pm 0.18.
\end{align}

Finally, we
estimate the combined contribution from the $a_{0}(980)$ and $f_{0}(980)$ mesons as
\begin{align}
a_{\mu }^{\mathrm{LbL,a_{0}+f_{0}}}\approx 0.01, 
\end{align}
and add it to the total result.

\section{Comparison with other models}

It should be mentioned that there are estimates of the scalar meson exchange contributions to the muon AMM in different versions of the local NJL model.
In \cite{Bijnens:1995xf} the combined scalar contribution was estimated as
\begin{equation}
a_{\mu }^{\mathrm{LbL,S}}=-(0.68 \pm 0.2) \cdot 10^{-10}\label{Bij},
\end{equation}
whereas \cite{Bartos:2001pg} gives the estimates for $\sigma$ and $a_0$ contributions as
\begin{align}
a_{\mu }^{\mathrm{LbL,\sigma}}&=(1.167\pm0.238) \cdot 10^{-10}, \nonumber \\
a_{\mu }^{\mathrm{LbL,a_0}}&=(0.062 \pm 0.024)\cdot 10^{-10}\label{BDKZ}.
\end{align}
One can see that our results (Table 1) are smaller in absolute values than other estimates.

Note, that in estimations (\ref{Bij}) and (\ref{BDKZ}) there is an ambiguity in the sign for the scalar meson exchange contributions.
In \cite{Blokland:2001pb} the analytical expressions for the pion and $\sigma$-meson contributions\footnote{For the pion exchange contribution, the coefficient of the leading, $\log^2(M_\rho/m_\mu)$, term in $(m_\mu/M_\rho)^2$ expansion was found in \cite{Knecht:2001qg}. } was obtained with the meson transition form factors taken from the simple vector meson dominance (VMD) model parameterized by the $\rho$-meson mass $M_\rho$. These expressions are given as an expansion in small parameters chosen in accordance with the mass scale hierarchy $({M}_{M}^2-m_\mu^2) \ll m_\mu^2 \ll M_\rho^2$. We reproduce numerically the coefficients of these expansions for the pseudoscalar meson contribution (Eqs.(8) and (10) of \cite{Blokland:2001pb}) \footnote{Namely, varying the mass parameter values for the $\rho$-meson, muon and difference between the pion and muon masses squared, one can extract different terms of the expansions given in \cite{Blokland:2001pb}.}
by using our code for numerical calculations of $a_\mu$ and substituting the VMD transition form factors instead of N$\chi $QM ones
\begin{align}
\mathrm{A}_{S}\left(  p^{2};q_{1}^{2},q_{2}^{2}\right)=\mathrm{F}^{VMD}_{M}\left( q_{1}^{2},q_{2}^{2}\right),\quad
\mathrm{B}_{S}\left(  p^{2};q_{1}^{2},q_{2}^{2}\right)=0.\nonumber
\end{align}
However, for the $\sigma$-meson contribution by using the same VMD model as in \cite{Blokland:2001pb},
we reproduce numerically the coefficients of the expansion in Eq. (11) of \cite{Blokland:2001pb} up to the overall sign.
Thus we conclude (as in \cite{Bartos:2001pg}) that the scalar meson contribution to the muon AMM for the VMD model has a positive sign in variance with the result of \cite{Blokland:2001pb}.

For additional check of computer code correctness, as suggested in \cite{Blokland:2001pb},
one can calculate the contribution to  the muon AMM from the vacuum polarization processes, $a_\mu^{\mathrm{HVP},M\gamma}$, 
where the virtual photon splits to the meson $M$ and the real photon $\gamma$.
As argued in \cite{Blokland:2001pb} these contributions have to be positive, since they are related by dispersion relations to the cross sections $\sigma(e^+e^-\to M\gamma)$. Our numerical results for the VMD model and a set of parameters used in \cite{Blokland:2001pb} (for $M_\sigma = M_\pi$) are
\begin{align}
a_{\mu,\mathrm{VMD} }^{\mathrm{LbL,}\pi^0}&=+5.64 \cdot 10^{-10}, \label{BCM1}\\
a_{\mu,\mathrm{VMD} }^{\mathrm{LbL,}\sigma}&=+4.76\cdot 10^{-10},\label{BCM2}\\
a_{\mu,\mathrm{VMD} }^{\mathrm{HVP,}\pi^0\gamma}&=+0.368 \cdot 10^{-10}, \label{BCM3}\\
a_{\mu,\mathrm{VMD} }^{\mathrm{HVP},\sigma\gamma}&=+0.265\cdot 10^{-10}.\label{BCM4}
\end{align}
Our numerical results for the pion exchange contributions, (\ref{BCM1}) and (\ref{BCM3}), agree with the results given in Eqs. (2) and (10) of \cite{Blokland:2001pb}.

In order to study the transition from the nonlocal model to the local one, we consider\footnote{Similar consideration was used in \cite{Radzhabov:2010dd} for the investigation of $1/N_c$ corrections.}
the nonlocal $SU(2)\times SU(2)$ model with Pauli--Villars regularization parameterized by
\begin{enumerate}
\item parameter of nonlocality $\Lambda$,
\item parameter of quark loop regularization $\Lambda_q$.
\end{enumerate}
The local model corresponds to the limit
\begin{eqnarray}
\Lambda\rightarrow\infty~,\quad
\Lambda_q^{}=\Lambda_q^{\rm fit}~,\quad
\label{localLimit}
\end{eqnarray}
while the nonlocal model without regularization can be obtained by setting
\begin{eqnarray}
\Lambda=\Lambda^{\rm fit}~,\quad
\Lambda_q^{}\rightarrow\infty~.\quad
\label{nonlocalLimit}
\end{eqnarray}
For definiteness, let us compare\footnote{We rescale current quark masses in order to reproduce mass of neutral pion instead of charged one.}
the local model\footnote{$m^{\mathrm{NJL}}_{c,u}=5.69$ MeV, $m^{\mathrm{NJL}}_{d,u}=253.9$ MeV, $\Lambda^{\mathrm{NJL}}_q=800$ MeV.}
\cite{Oertel:1999fk,Oertel:2000jp}  with the nonlocal
one\footnote{$m^{\mathrm{ N\chi QM}}_{c,u}=5.45$ MeV, $m^{\mathrm{ N\chi QM}}_{d,u}=255.8$ MeV, $\Lambda^{\mathrm{ N\chi QM}}=902.4$ MeV.}
\cite{Blaschke:2007np}.
The values of the quark condensate in the local and the nonlocal models coincide numerically within less than $0.5~\%$ deviation.

\begin{figure}[t]
\resizebox{0.45\textwidth}{!}{  \includegraphics{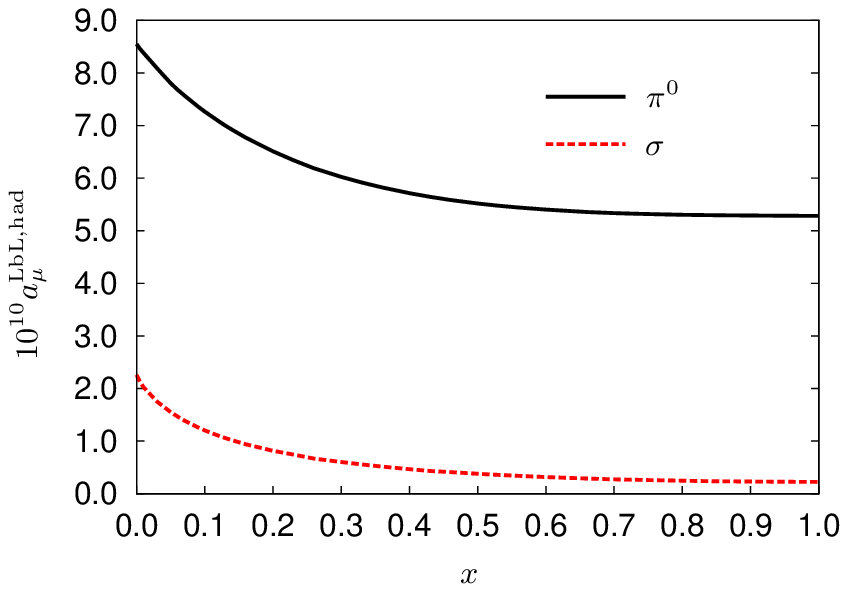}
}
\caption{ LbL contribution to the muon AMM
from the neutral pion and $\protect\sigma $ exchanges in nonlocal model with additional regularization. Zero $x$ corresponds to a local NJL model while $x$ equal 1 is the nonlocal model without regularization. }
\label{fig:PiSiLbLNJL}
\end{figure}

The result is presented in Fig. \ref{fig:PiSiLbLNJL}, where we introduce the parameter of nonlocality $x$.
Zero of $x$ corresponds to the local NJL model, while the nonlocal model is reproduced for $x$ equal to one. For each $x$ point in between zero and one the regularization parameter behaves as $\Lambda_q=\Lambda^{\mathrm{NJL}}_q/(1-x)$, the dynamical quark mass scales linearly and the current quark mass and $\Lambda$ are refitted in order to reproduce the mass of the neutral pion and linearly scaled quark condensate.

One can see that the pion contribution to the muon AMM increases from $5.3\cdot 10^{-10}$ in the N$\chi $QM model to $8.5\cdot 10^{-10}$ in the local NJL model.
More dramatic situation takes place for the $\sigma$-meson contribution. The contribution in the local limit is ten times larger than in the nonlocal model ($2.2\cdot 10^{-10}$ instead of $0.22\cdot 10^{-10}$).
The values obtained in the local limit are of the same order as numbers quoted in \cite{Bartos:2001pg}.

\section{Conclusions}

We found that within the N$\chi $QM the pseudoscalar meson contributions to
muon AMM are systematically lower then the results
obtained in the other works (see discussion in \cite{Dorokhov:2011zf}). The
full kinematic dependence of the vertices on the pion virtuality diminishes
the result by about 20-30\% as compared to the case where this dependence is
neglected. For $\eta $ and $\eta ^{\prime }$ mesons the results are reduced
by factor about 3 in comparison with the results obtained in other models
where the kinematic dependence was neglected (see details in \cite%
{Dorokhov:2011zf}). The total contribution of pseudoscalar exchanges
\begin{equation}
a_{\mu }^{\mathrm{LbL,PS}}=(5.85\pm 0.87)\cdot 10^{-10}
\end{equation}%
is approximately by factor 1.5 less than the most of previous estimates.

The scalar mesons
contribution is positive and partially cancels model dependence of the
pseudoscalar contribution. The combined value for the
scalar--pseudoscalar contribution is estimated as
\begin{equation}
a_{\mu }^{\mathrm{LbL,PS+S}}=(6.25\pm 0.83)\cdot 10^{-10}.
\end{equation}

It is well known that the hadronic LbL contribution to the muon AMM calculated in the effective approaches have model dependent features. However, there are few constraints following from QCD and need to be satisfied in any acceptable model calculations. One result was obtained by Melnikov and Vainshtein \cite{Melnikov:2003xd} (see also for discussions \cite{Prades:2009tw}) for the high photon momentum behavior of the total LbL amplitude. The consistency of the N$\chi $QM with the Melnikov-Vainshtein constraints was carefully analyzed in our previous works \cite{Dorokhov:2008pw,Dorokhov:2011zf}. It is based on the fact that in the high photon momentum limit the dynamical nonperturbative dressing effects containing in the quark box diagram becomes vanishing and this diagram produces short range behavior characteristic for the perturbative QCD regime. At the same time the LbL contributions containing the meson exchanges are responsible for the long distance dynamics and are suppressed in the high photon momentum limit.

Another important constraint concerns the low photon momentum limit of the LbL contribution with intermediate pion exchange. In \cite{Knecht:2001qg} (see also for discussions \cite{Prades:2009tw}) the coefficient of the leading
logarithm of ultraviolet regulator, 
$\log^2\Lambda$,
arising in this contribution was computed in the chiral limit in the leading in $1/N_c$ approximation.
In \cite{Knecht:2001qg} there was also discussed the correspondence between this result and the model calculations given in \cite{Knecht:2001qf}. They found that for the case of the vector meson dominance form factor and in the limit $M_V\to\infty$ the logarithmic coefficient numerically agrees with chiral perturbative theory result.
We have checked this statement by performing the chiral expansion of the polarization operators as well as triangle diagram functions. Then, one can indeed reproduce the correct coefficients in front of $\log^2\Lambda$ terms in the limit as the ultraviolet regulator $\Lambda$ goes to infinity.

Finally, the important point for the model calculations is the total contribution from all leading diagrams. This is because the different models may redistribute partial contributions differently. In the present work we show that the contribution of the scalar mesons being relatively small leads to stabilization of the total pseudoscalar and scalar contribution with respect to variation of the model parameters. The next step is the calculation of the dynamical quark and pion loops contribution which is now in progress.

We thank M.~Buballa, Yu.~M.~Bystritskiy, C.~Fischer, N.~I.~Kochelev,  E.~A.~Kuraev,  V.~P.~Lomov, B.-J.~Schaefer, and R.~Williams for critical remarks
and illuminating discussions.
A.E.D. and A.E.R. are grateful for the hospitality during visits at the TU Darmstadt.

This work is supported in part by the Heisenberg-Landau program (JINR),
the Russian Foundation for Basic Research (projects No.~10-02-00368 and No.~11-02-00112),
the Federal Target Program Research and Training Specialists
in Innovative Russia 2009-2013 (16. 740.11.0154, 14.B37.21.0910).

\appendix
\section{Four-quark coupling constants, polarization operators and mixing angles}
\label{AppA}

The elements of $\mathbf{G}_{ch}$-matrices take the form
\begin{align}
&G_{00}  = G \pm \frac{H}{3} (2S_u  + S_s ), \quad% \\
G_{88}  = G \mp \frac{H}{6} (4S_u  - S_s ), \nonumber\\
&G_{08}  = G%^{S,P}
_{80}= \mp\frac{\sqrt{2}}{6}H(S_u -S_s ) , \quad%\nonumber\\
%&G_{(a_0,\pi^0)} =
G_{33}=G \mp\frac{H}{2} S_s, \label{GConts}%\nonumber
\end{align}
where the upper sign corresponds to the scalar channel, while the lower sign corresponds to the pseudoscalar channel. For the pion, $G_{\pi}$ equals to $G_{33}$ of the pseudoscalar interaction, and, for the $a_0$-meson, $G_{a_0}$ equals to $G_{33}$ of the scalar interaction.

The elements of $\mathrm{\Pi}_{ch}(P^{2})$-matrix for the scalar and pseudoscalar mesons are diagonal in the quark-flavor basis, and in
the singlet-triplet-octet basis they are given by
\begin{align}
&\Pi_{00}(P^{2})  = \frac{1}{3}(2\Pi_{uu}(P^{2}) +\Pi_{ss}(P^{2}) ),\nonumber\\
&\Pi_{88}(P^{2})  = \frac{1}{3}(\Pi_{uu}(P^{2}) +2\Pi_{ss}(P^{2}) ),\\
&\Pi_{08}(P^{2})  = \Pi_{80}(P^{2}) =\frac{\sqrt{2}}{3}(\Pi_{uu}(P^{2}) -\Pi_{ss}(P^{2}) ),\nonumber\\
%&\Pi_{(a_0,\pi^0)}(P^{2})=
&\Pi_{33}(P^{2})=\Pi_{uu}(P^{2}),\nonumber
\end{align}
where the difference between the scalar and pseudoscalar channels is in the polarization operators
\begin{align}
\Pi_{ij}(P^{2})=&8N_{c}\int\frac{d_{E}^{4}K}{(2\pi)^{4}}\frac{f^{2}(K_{+}^{2})f^{2}(K_{-}^{2}) }{D_{i}(K_{+}^{2})D_{j}(K_{-}^{2}) }\times\nonumber\\
&\times\left[  (K_{+}\cdot K_{-}) \mp m_{i}(K_{+}^{2})m_{j}%
(K_{-}^{2})\right] ,
\end{align}
where $K_{\pm}=K\pm P/2$. Similarly to Eq. (\ref{GConts}), the upper sign corresponds to the scalar channel and the lower sign corresponds to the pseudoscalar channel, $\Pi_{a_0}$ equals to $\Pi_{33}$ for the scalar channel and $\Pi_{\pi}$ equals to $\Pi_{33}$ the pseudoscalar channel. The unrenormalized mesonic propagators for the
scalar mesons are
\begin{align}
&D_{a_0}^{-1}(P^{2})=-G_{a_0}^{-1}+\Pi_{a_0}(P^{2}),\nonumber\\
&D_{\sigma,f_0}^{-1}(P^{2})=\frac{1}{2}\left[  (A+C) \pm \sqrt{(A-C)^{2}+4B^{2}}\right]  ,\nonumber\\
 &A=-G_{88}/\mathrm{det}(\mathbf{G}_{ch})+\Pi_{00}(P^{2}), \label{UnRenormPropag}\\
 &B=+G_{08} /\mathrm{det}(\mathbf{G}_{ch})+\Pi_{08}(P^{2}),\nonumber\\
 &C=-G_{00}/\mathrm{det}(\mathbf{G}_{ch})+\Pi_{88}(P^{2}),\nonumber\\
&\mathrm{det}(\mathbf{G}_{ch})=G_{00}G_{88}-G_{08}^{2}.\nonumber
\end{align}
The mixing angle depends on the meson virtuality
\begin{align}
\theta_{S}(P^{2})=\frac{1}{2}\arctan\left[  \frac{2B}%
{A-C}\right]  -\frac{\pi}{2}\Theta\left(  A-C\right).
\label{AngelMix}
\end{align}

Expressions for the unrenormalized propagators for the pseudoscalar mesons are similar to the scalar meson propagators, Eqs. (\ref{UnRenormPropag}),  (\ref{AngelMix}), with replacements $a_0 \to \pi$,  $\sigma \to \eta$,  $f_0 \to \eta^\prime$ and $\theta_{S} \to \theta_{P}$.

\section{Feynman rules for nonlocal vertices}
\label{AppB}
\begin{figure}[h]
\begin{tabular*}{\columnwidth}{@{\extracolsep{\fill}}cccc@{}}
\resizebox{0.15\textwidth}{!}{\includegraphics{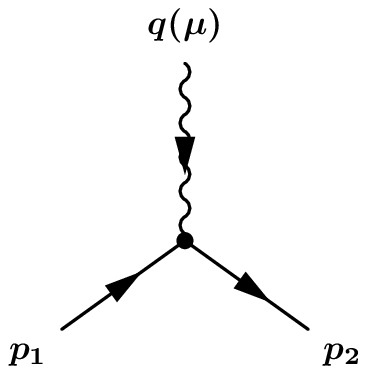}} &
\resizebox{0.15\textwidth}{!}{\includegraphics{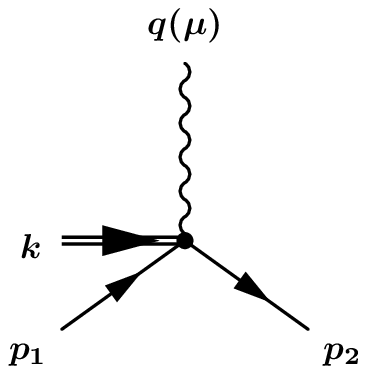}} \\ %&
%\resizebox{0.11\textwidth}{!}{\includegraphics{G3}} &
%\resizebox{0.11\textwidth}{!}{\includegraphics{G4}}    \\
(a)&(b) %&(c)&(d)
\end{tabular*}
\caption{Vertices $\mathrm{\Gamma}^\mu_{p_2,p_1}$, Eq. (\ref{GammaMu}), $\mathrm{\Gamma}^{M;\mu}_{p_2,p_1,q}$, Eq. (\ref{GammaMppq}), with one photon. }
\label{fig:Vertices1}
\end{figure}

The total vertex of photon interaction with quark-antiquark pair (Fig. \ref{fig:Vertices1}a) contains local and nonlocal parts
\begin{align}
\mathrm{\Gamma}^\mu_{p_2,p_1}=T^a(\gamma_\mu-(p_{1}+p_{2})_{\mu}\mathrm{m}^{(1)}(p_1,p_2)), \label{GammaMu}
\end{align}
where $T^a \equiv Q$ and $\mathrm{m}^{(1)}(p_1,p_2)$ is the first order finite-difference of the dynamical quark mass
\begin{align}
\mathrm{m}^{(1)}(p_1,p_2) = \frac{m_{i}(p_1^{2}) - m_{i}(p_2^{2})}{p_{1}^2-p_{2}^2}.
\end{align}

The contact interaction vertex of meson, photon and quark-anti\-quark pair (Fig.\ref{fig:Vertices1}b) is purely nonlocal and takes the form
\begin{align}
\mathrm{\Gamma}^{M;\mu}_{p_2,p_1,q}=& -g_{M}(k)\left(\mathrm{f}^{(1)}(p_{1},p_{1}+q) f(p_{2})(2p_{1}+q)_{\mu}T^a\Gamma _{M}^{b}+\right.\nonumber\\
&\left.+\mathrm{f}^{(1)}(p_{2},p_{2}-q)f(p_{1})(2p_{2}-q)_{\mu}\Gamma _{M}^{b}T^a\right) . \label{GammaMppq}
\end{align}

\begin{figure}[h]
\begin{tabular*}{\columnwidth}{@{\extracolsep{\fill}}cccc@{}}
\resizebox{0.15\textwidth}{!}{\includegraphics{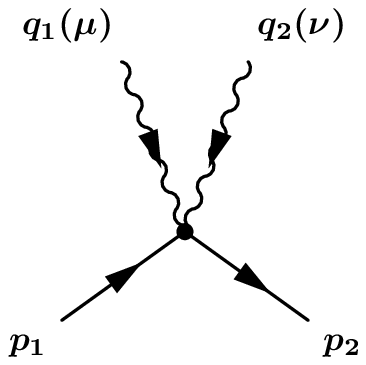}} &
\resizebox{0.15\textwidth}{!}{\includegraphics{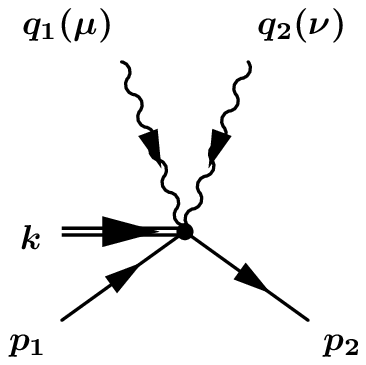}}    \\
(a)&(b) %&(c)&(d)
\end{tabular*}
\caption{Vertices $\mathrm{\Gamma}^{\mu,\nu}_{p_2,p_1,q_1,q_2}$, Eq. (\ref{Gammappqq}), and  $\mathrm{\Gamma}^{M;\mu,\nu}_{p_2,p_1,q_1,q_2}$, Eq. (\ref{GammaMppqq}),  with two photons.}
\label{fig:Vertices2}
\end{figure}

In order to express the vertices with two external photons we introduce
the following functions
\begin{align}
&G_{\mu}^{a}\left(k,q\right) =iT^{a}\left( 2k+q\right) _{\mu}\mathrm{f}^{(1)}(k,k+q),\nonumber\\
&G_{\mu \nu }^{ab}\left( k,q,q^{\prime },k^{\prime }\right) =
-f\left( k^{\prime }\right) \biggl\{
T^{a}T^{b}\left[ g_{\mu \nu}\mathrm{f}^{(1)}(k,k+q+q^{\prime })+\right. \biggr. \nonumber\\
&+\left. \left.\left[ 2\left( k+q^{\prime }\right) +q\right] _{\mu }\left( 2k+q^{\prime }\right) _{\nu }\mathrm{f}^{(2)}\left( k,k+q^{\prime },k+q+q^{\prime }\right) \right] \right. +\nonumber\\
&\quad+\biggl.\left[ \left( q,a,\mu \right) \longleftrightarrow \left( q^{\prime },b,\nu \right) \right] \biggr\} ,
\end{align}
where $\mathrm{f}^{(2)}\left( k_1,k_2,k_3\right)$ is the second order finite-difference
\begin{align}
\mathrm{f}^{(2)}\left( k_1,k_2,k_3\right) = \frac{\mathrm{f}^{(1)}(k_1,k_3)-\mathrm{f}^{(1)}(k_1,k_2)}{k_3^2-k_2^2}.
\end{align}
With this notation, the vertex of two-photon interaction with quark-anti\-quark pair (Fig.\ref{fig:Vertices2}a) is
\begin{align}
&\mathrm{\Gamma}^{\mu,\nu}_{p_2,p_1,q_1,q_2}= m_d\biggl\{ G_{\mu \nu }^{ab}\left( p_1,q_1,q_2,p_2 \right) + \nonumber \\
&+G_{\mu \nu }^{ab}\left(
p_2-q_1-q_2,q_1,q_2,p_1\right)\biggr.  +\label{Gammappqq} \\
&\biggl. + G_{\mu }^{a}\left( p_1,q_1\right) G_{\nu }^{b}\left( p_2-q_2,q_2\right) + G_{\mu }^{a}\left(
p_2-q_1,q_1\right) G_{\nu }^{b}\left( p_1,q_2\right) \biggr\},\nonumber
\end{align}
and the interaction vertex for two photons, meson and quark-anti\-quark pair (Fig.\ref{fig:Vertices2}b) becomes
\begin{align}
\mathrm{\Gamma}^{M;\mu,\nu}_{p_2,p_1,q_1,q_2}=&-g_{M}(k)\biggl\{ G_{\mu \nu }^{ab}\left( p_1,q_1,q_2,p_2 \right)\Gamma_{M}^c + \nonumber \\
&+\Gamma_{M}^c G_{\mu \nu }^{ab}\left(
p_2-q_1-q_2,q_1,q_2,p_1\right)\biggr.  \nonumber \\
&\biggl. + G_{\mu }^{a}\left( p_1,q_1\right)\Gamma_{M}^c G_{\nu }^{b}\left( p_2-q_2,q_2\right) \label{GammaMppqq}\\
&+ G_{\mu }^{a}\left(
p_2-q_1,q_1\right) \Gamma_{M}^c G_{\nu }^{b}\left( p_1,q_2\right) \biggr\}.\nonumber
\end{align}

\section{Amplitude with meson and two photons}
\label{AppC}
\begin{figure*}[t]
\begin{tabular*}{\textwidth}{@{}ccccccccccccc@{}}
\raisebox{-0.5\height}{\resizebox{!}{0.09\textheight}{\includegraphics{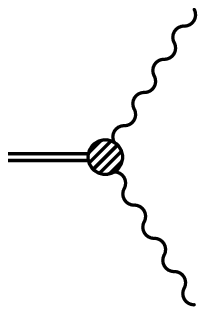}}}&\raisebox{-0.5\height}{=}& %&
\raisebox{-0.5\height}{\resizebox{!}{0.09\textheight}{\includegraphics{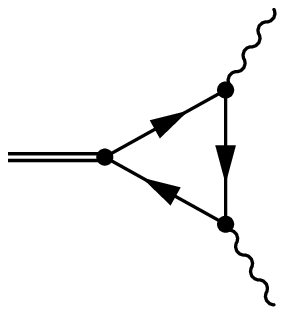}}}&\raisebox{-0.5\height}{+}& %&
\raisebox{-0.5\height}{\resizebox{!}{0.09\textheight}{\includegraphics{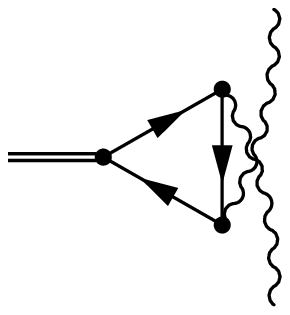}}}&\raisebox{-0.5\height}{+}& %&
\raisebox{-0.5\height}{\resizebox{!}{0.09\textheight}{\includegraphics{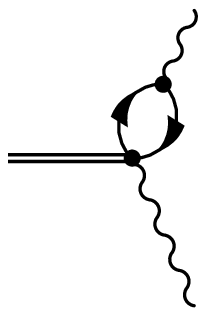}}}&\raisebox{-0.5\height}{+}& %\\ %&
\raisebox{-0.5\height}{\resizebox{!}{0.09\textheight}{\includegraphics{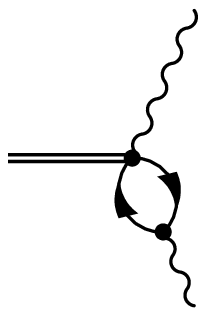}}}&\raisebox{-0.5\height}{+}& %\\ %&
\raisebox{-0.5\height}{\resizebox{!}{0.09\textheight}{\includegraphics{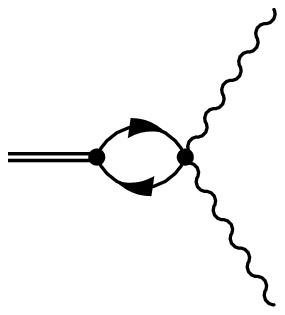}}}&\raisebox{-0.5\height}{+}& %\\ %&
\raisebox{-0.5\height}{\resizebox{!}{0.09\textheight}{\includegraphics{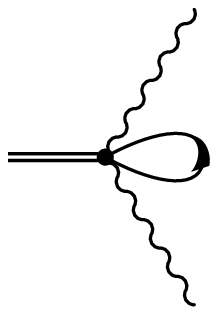}}}\\
(a)&&(b)&&(c)&&(d)&&(e)&&(f)&&(g)
\end{tabular*}
\caption{The diagrams for the photon-meson transition.}
\label{fig:SggSimp}
\end{figure*}

The photon-meson transition amplitude is a sum of diagrams shown in Fig. \ref{fig:SggSimp}, where all particles are virtual. For the scalar meson it takes the form
\begin{align}
A&\left(\gamma_{1}^{\ast}\gamma_{2}^{\ast}\rightarrow M^{\ast}\right)
 =e^{2}\epsilon_{1}^{\mu}\epsilon_{2}^{\nu}\Delta^{\mu\nu}\left( p, q_{1}, q_{2}\right), \nonumber\\
\Delta&^{\mu\nu}\left( p, q_{1}, q_{2}\right)= -i N_{c}
\int\frac{d^{4}k}{(2\pi)^{4}} \label{DeltaVertex} \\
&\quad\times\mathrm{Tr} \bigg(2 \mathrm{\Gamma}^{M}_{k_2,k_1}S(k_1)\mathrm{\Gamma}^\mu_{k_1,k_3}S(k_3)\mathrm{\Gamma}^\nu_{k_3,k_2}S(k_2)\nonumber\\
&\quad+ \mathrm{\Gamma}^{M;\mu}_{k_2,k_3,q_1}S(k_3)\mathrm{\Gamma}^\nu_{k_3,k_2}S(k_2)
 + \mathrm{\Gamma}^{M;\nu}_{k_3,k_1,q_2}S(k_1)\mathrm{\Gamma}^\mu_{k_1,k_3}S(k_3) \nonumber\\
&\quad+ \mathrm{\Gamma}^{M}_{k_2,k_1}S(k_1)\mathrm{\Gamma}^{\mu,\nu}_{k_1,k_2,q_1,q_2}
+ \mathrm{\Gamma}^{M;\mu,\nu}_{k_3,k_3,q_1,q_2}S(k_3) \bigg),\nonumber
\end{align}
where the symbols are the photon momenta $q_{1,2}$, the photon polarization vectors $\epsilon _{1,2}$, the meson momentum $p=q_{1}+q_{2}$, and the quark momenta $k_{1,2,3}$ ($k_{1}=k+q_{1}$, $k_{2}=k-q_{2}$, $k_3=k$). The first term in parentheses corresponds to the quark triangle diagrams\footnote{In the case of pseudoscalar mesons, the diagrams in Fig. \ref{fig:SggSimp}d-g give a zero contribution due to chirality considerations.
}
(Fig. \ref{fig:SggSimp}b and crossed term Fig. \ref{fig:SggSimp}c) and next terms corresponds to the diagrams in Figs. \ref{fig:SggSimp}d-g with effective nonlocal vertices defined in  (\ref{GammaMppq}), (\ref{Gammappqq}), (\ref{GammaMppqq}).

For different scalar meson states one has the following combinations of nonstrange and strange components 
\begin{align}
&\Delta^{\mu\nu}_{a_{0}}\left(  p,q_{1},q_{2}\right)
   =g_{a_0}(p^{2})\delta^{\mu\nu}_{u}\left(  p^{2};q_{1}^{2} ,q_{2}^{2}\right)  ,\nonumber\\
&\Delta^{\mu\nu}_{\sigma}\left(  p,q_{1},q_{2}\right)
   =\frac{g_{\sigma}(p^{2})}{3\sqrt{3}}\times\nonumber\\
&\quad\times\biggl[  \left(5\delta^{\mu\nu}_{u}\left(  p^{2};q_{1}^{2},q_{2}^{2}\right)  -2\delta^{\mu\nu}_{s}\left(  p^{2};q_{1}^{2},q_{2}^{2}\right)  \right) \cos\theta_S(p^{2})-  \nonumber\\
&\quad\, -\sqrt{2}\left(  5\delta^{\mu\nu}_{u}\left(  p^{2};q_{1}^{2},q_{2}^{2}\right)  +\delta^{\mu\nu}_{s}\left(  p^{2};q_{1}^{2},q_{2}^{2}\right)  \right)\sin\theta_S(p^{2})\biggr]  ,\nonumber\\
&\Delta^{\mu\nu}_{f_0}\left(  p,q_{1},q_{2}\right)  =\frac{g_{f_0}(p^{2})}{3\sqrt{3}} \times\\
&\quad\times\biggl[ \left(  5\delta^{\mu\nu}_{u}\left(  p^{2};q_{1}^{2},q_{2}^{2}\right)-2\delta^{\mu\nu}_{s}\left(  p^{2};q_{1}^{2},q_{2}^{2}\right)  \right)  \sin\theta_S(p^{2})+  \nonumber\\
& \quad\, +\sqrt{2}\left(  5\delta^{\mu\nu}_{u}\left(  p^{2};q_{1}^{2},q_{2}^{2}\right)  +\delta^{\mu\nu}_{s}\left(  p^{2};q_{1}^{2},q_{2}^{2}\right)  \right) \cos\theta_S(p^{2}) \biggr].\nonumber
\end{align}

One can easily see from Eqs. (\ref{ApiGG}), (\ref{BPrime}) that the mixing for the form-factors $\mathrm{A}_{S}$, $\mathrm{B}_{S}$, $\mathrm{B}^\prime_{S}$ from the components $A_{u}$, $B_{u}$, $B^\prime_{u}$ and $A_s$, $B_{s}$, $B^\prime_{s}$ is similar.
One should project $A_{i}$ and $B_{i}$  ($i=u,s$) from loops of nonstrange and strange quarks
\begin{align}
A_i\left(  p^{2};q_{1}^{2} ,q_{2}^{2}\right)&=\frac{\delta_i^{\mu\nu}\left( p^2, q_{1}^2, q_{2}^2\right) }{2(q_1\cdot q_2)}\left[ g^{\mu\nu}-\frac{(q_1\cdot q_2)\;q_1^\nu q_2^\mu}{(q_1\cdot q_2)^2-q_1^2 q_2^2}\right], \nonumber\\
B_i\left(  p^{2};q_{1}^{2} ,q_{2}^{2}\right)&=-\frac{\delta_i^{\mu\nu}\left( p^2, q_{1}^2, q_{2}^2\right) }{2(q_1\cdot q_2)\left((q_1\cdot q_2)^2-q_1^2 q_2^2\right)}\times  \nonumber\\
&\times \left[ g^{\mu\nu}-3\frac{(q_1\cdot q_2)\;q_1^\nu q_2^\mu}{(q_1\cdot q_2)^2-q_1^2 q_2^2}\right], \label{NonlocAB}
\end{align}
where
\begin{align}
\delta_i^{\mu\nu}\left( p^2, q_{1}^2, q_{2}^2\right)
 =&- 2i
\int\frac{d^{4}k}{(2\pi)^{4}}
\left[J^{\mu\nu}_{bc}+J^{\mu\nu}_{de}+J^{\mu\nu}_f+J^{\mu\nu}_g\right],\nonumber
\end{align}
and different terms in brackets, $J^{\mu\nu}$, correspond to the diagrams shown in Figs. \ref{fig:SggSimp}, with lower indices being the symbol of the figure.

Below for simplicity, a momentum is denoted as a lower index and a quark flavor index $i$ is omitted
\begin{align}
&f_n \equiv  f(k_n^2), \; m_n \equiv m_i(k_n^2),\; D_n \equiv D_i(k_n^2),\; f^{(1)}_{nm}\equiv \mathrm{f}^{(1)}(k_n,k_m), \nonumber \\
&\mathrm{m}^{(1)}_{ln} \equiv \mathrm{m}^{(1)}(k_n,k_m),\; \mathrm{f}^{(2)}_{nml} \equiv \mathrm{f}^{(2)}\left( k_n,k_m,k_l\right).\nonumber %\\
\end{align}

Then, one has
\begin{align}
&J^{\mu\nu}_\mathrm{bc}=\frac{f_1f_2}{D_1D_2D_3}\biggl[
V_1^{\mu\nu}+\mathrm{m}^{(1)}_{23}\mathrm{m}^{(1)}_{13}(k_2+k_3)^{\nu}(k_1+k_3)^{\mu}V_4-\nonumber\\
&\quad\quad-\mathrm{m}^{(1)}_{23}(k_2+k_3)^{\nu}V_2^{\mu}-\mathrm{m}^{(1)}_{13}(k_1+k_3)^{\mu}V_3^{\nu}
\biggr], \label{Jbc} %\nonumber
\end{align}
where
\begin{align}
V_1^{\mu\nu}&=m_{1}\left[k_{2}^{\nu}k_3^{\mu}+k_{2}^{\mu}k_3^{\nu}\right]
+m_3\left[k_{2}^{\nu}k_{1}^{\mu}-k_{2}^{\mu}k_{1}^{\nu}\right]+\nonumber\\
&+m_{2}\left[k_3^{\nu}k_{1}^{\mu}+k_3^{\mu}k_{1}^{\nu}\right]+g^{\mu\nu}\left[m_{1}m_3m_{2}-m_{1}(k_{2}k_3)+\right.\nonumber\\
&+\left.m_3(k_{2}k_{1})-m_{2}(k_3k_{1})\right],\nonumber
\\
V_2^{\mu}&=
  k_1^{\mu}\left[m_2m_3+(k_2k_3)\right]
 +k_2^{\mu}\left[m_1m_3-(k_1k_3)\right]+\nonumber\\
&+k_3^{\mu}\left[m_1m_2+(k_1k_2)\right],\nonumber
\\
V_3^{\nu}&=
  k_1^{\nu}\left[m_2m_3-(k_2k_3)\right]
 +k_2^{\nu}\left[m_1m_3+(k_1k_3)\right]+\nonumber\\
&+k_3^{\nu}\left[m_1m_2+(k_1k_2)\right],
\nonumber
\\
V_4&=m_1m_2m_3+m_1(k_2k_3)+m_3(k_1k_2)+m_2(k_1k_3),\nonumber %\Bigg\right]
\end{align}
and
\begin{align}
&J^{\mu\nu}_\mathrm{de}=\frac{f_2}{D_2D_3}\mathrm{f}^{(1)}_{13}(k_1+k_3)^{\mu}\times\nonumber\\
&\quad\times\biggl(m_2k_3^{\nu} + m_3k_2^{\nu} -\mathrm{m}^{(1)}_{32}(k_2+k_3)^{\nu}((k_3k_2)+m_3m_2)\biggr)+\nonumber\\
&\quad+\frac{f_1}{D_1D_3}\mathrm{f}^{(1)}_{23}(k_2+k_3)^{\nu}\times \label{Jde} \\ %\nonumber\\
&\quad\times\biggl(m_3k_1^{\mu} +m_1k_3^{\mu}-\mathrm{m}^{(1)}_{13}(k_1+k_3)^{\mu}((k_3k_1)+m_3m_1)\biggr),\nonumber
\end{align}
\begin{align}
&J^{\mu\nu}_\mathrm{f}=\frac{f_1f_2}{D_1D_2}((k_1k_2)+m_1m_2)m_d
\bigg[(f_1+f_2) g^{\mu\nu}\mathrm{f}^{(1)}_{12}+ \nonumber \\
&\,\;+(k_1+k_3)^{\mu}(k_2+k_3)^{\nu}\left((f_1+f_2)(\mathrm{f}^{(2)}_{231}+\mathrm{f}^{(2)}_{132})-\mathrm{f}^{(1)}_{13}\mathrm{f}^{(1)}_{23}\right)\bigg],\nonumber
\end{align}

\begin{align}
&J^{\mu\nu}_\mathrm{g}=
     -\frac{f_2m_2}{D_2}  [g^{\mu\nu}\mathrm{f}^{(1)}_{12}+(k_1+k_3)^{\mu}(k_2+k_3)^{\nu}\mathrm{f}^{(2)}_{231}]  -\nonumber\\
&\quad-\frac{f_1m_1}{D_1}  [g^{\mu\nu}\mathrm{f}^{(1)}_{12}+(k_2+k_3)^{\nu}(k_1+k_3)^{\mu}\mathrm{f}^{(2)}_{132}]  +\nonumber\\
&\quad+\frac{f_3m_3}{D_3}  \mathrm{f}^{(1)}_{13}\mathrm{f}^{(1)}_{23}(2k_1+k_3)^{\mu}(2k_2+k_3)^{\nu}.\nonumber
\end{align}

Analytical expressions for the form factors $A_i\left(  p^{2};q_{1}^{2} ,q_{2}^{2}\right)$ and $B_i^\prime\left(  p^{2};q_{1}^{2} ,q_{2}^{2}\right)$ in the case of special kinematics\footnote{$B_i\left(  p^{2};q_{1}^{2} ,q_{2}^{2}\right)$ is divergent in this kinematic.}, when
one photon is real, $q_1^2=0$, and the virtuality of second photon is equal to the virtuality of meson $p_2^2=q_2^2$, can be obtained by expanding the quark-loop expressions, Eqs. (\ref{NonlocAB}), (\ref{Jbc}), (\ref{Jde}), in $q_1^2$. The resulting expressions contain derivatives of the nonlocal function $f(k^2)$ up to third order. These expressions are rather cumbersome and not presented here. Alternatively, one can calculate the form factors for small but nonzero $q_1^2$ and then take the limit numerically.

\section{Local limit of  $\gamma^{\ast} \gamma^{\ast} \to S^\ast $ amplitude}
\label{AppE}
In the local model with constituent quark masses $m_i$ ,the triangle quark-loop diagrams, depicted in Figs. \ref{fig:SggSimp}b-c, reduce to the following expression
\begin{align}
&\delta_{\mathrm{loc};i}^{\mu\nu}\left( p^2, q_{1}^2, q_{2}^2\right)=m_i g^{\mu\nu}I_{\mathrm{g}}(m_i^2)+{A}_{\mathrm{loc};i}\left(p^{2};q_{1}^{2},q_{2}^{2}\right)\times
\nonumber\\
&\quad  \quad \times P_{A}^{\mu \nu }(q_{1},q_{2})+{B}_{\mathrm{loc};i}(p^{2};q_{1}^{2},q_{2}^{2})P_{B}^{\mu \nu }(q_{1},q_{2}),
\end{align}
where $I_{\mathrm{g}}(m_i^2)$ is a gauge non-invariant term (constant)
\begin{align}
&I_{\mathrm{g}}(m_i^2) =\frac{1}{2 \pi^2 } \int\limits^{1}_{0}dx_1\int\limits^{1-x_1}_{0} dx_2
\frac{m_i^2-X}{m_i^2-X}=\frac{1}{4 \pi^2 } , \\
&\quad\quad X = x_1 (1-x_1-x_2) q_2^2+x_2 (1-x_1-x_2) q_1^2+x_1 x_2 p^2, \nonumber
\end{align}
which should be eliminated by suitable regularization, e.g., the Pauli-Villars regularization $I_{\mathrm{g}}(m_i^2)-I_{\mathrm{g}}(\Lambda_\mathrm{PV}^2)=0$, and the form factors read
\begin{align}
&A_{\mathrm{loc};i}\left(  p^{2};q_{1}^{2} ,q_{2}^{2}\right)=\frac{m_i}{4 (q_1 q_2)\pi^2 } \int\limits^{1}_{0}dx_1\int\limits^{1-x_1}_{0} dx_2 \times\nonumber\\
&\quad\quad\times \frac{4X-p^2 +q_2^2 (1 -2x_1)+q_1^2 (1 -2x_2)}{m_i^2-X}, \\
&B_{\mathrm{loc};i}\left(  p^{2};q_{1}^{2} ,q_{2}^{2}\right)=\frac{m_i}{(q_1 q_2)\pi^2 } \int\limits^{1}_{0}dx_1\int\limits^{1-x_1}_{0} dx_2 \frac{ x_2 (1 -2  x_2)}{q_2^2(m_i^2-X)}.\nonumber
\end{align}
Note that, if one takes the local limit of the nonlocal expression Eq. (\ref{NonlocAB}) by setting $\Lambda \to \infty$, the contribution of nonlocal diagrams completely cancel a gauge noninvariant term.

For special kinematics considered above, the form-factors become $(\bar{x} = 1-x)$
\begin{align}
&A_{\mathrm{loc};i}\left(  p^{2}; p^{2} ,0\right)=-\frac{m_i}{12 \pi^2 } \int\limits^{1}_{0}dx 
\frac{2 m_i^2 - p^2  \bar{x} x\left(1+4x\bar{x}\right)}{(m_i^2-x\bar{x}p^2)^2}\, , \nonumber\\
&B^\prime_{\mathrm{loc};i}\left(  p^{2}; p^{2} ,0\right)=
-\frac{m_i}{6 \pi^2 }\int^{1}_{0}dx\frac{1-6x \bar{x}}{m_i^2-p^2x\bar{x}}
\,.
\end{align}

\section{Tensor structures for LbL amplitude}
\label{AppD}
Functions averaged over muon momenta can be represented as
\begin{align}
 \mathrm{Ts}^{\mathrm{XY}}_{\mathbf{i}} = \sum \limits _{j=1}^6 \left\langle A \right\rangle_j Z^{\mathrm{XY}}_{\mathbf{i},j},
\end{align}
where $\left\langle A \right\rangle_j$ are the averages of scalar products with muon momentum in the numerator and muon propagators in the denominator ($\mathrm{D}_1=\left(P+Q_{1}\right)^{2}+m_{\mu}^{2}$, $\mathrm{D}_2=\left(P-Q_{2}\right)^{2}+m_{\mu}^{2}$)
\begin{align}
&\left\langle A \right\rangle_1= \left\langle \frac{1}{\mathrm{D}_{1}}\right\rangle =\frac{R_{1}-1}{2m_{\mu}^{2}}\,,\quad
 \left\langle A \right\rangle_2= \left\langle \frac{1}{\mathrm{D}_{2}}\right\rangle =\frac{R_{2}-1}{2m_{\mu }^{2}}\,,\nonumber\\
& \left\langle A \right\rangle_3 =\left\langle \frac{PQ_{2}} {\mathrm{D}_{1}}\right\rangle =(Q_{1}Q_{2})\frac{\left(  1-R_{1}\right)  ^{2}}{8m_{\mu}^{2}}\,\,,\nonumber\\
& \left\langle A \right\rangle_4= \left\langle \frac{PQ_{1}}{\mathrm{D}_{2}}\right\rangle =-(Q_{1}Q_{2})\frac{\left( 1-R_{2}\right)  ^{2}}{8m_{\mu}^{2}}\,\,,\,\,\,\label{Not}\\
& \left\langle A \right\rangle_5= \left\langle \frac{1}{\mathrm{D}_{1}  \mathrm{D}_{2}}\right\rangle =\frac{1}{M_{\mu}^{2}\left\vert Q_{1}\right\vert \left\vert Q_{2}\right\vert x}\arctan\left[  \frac{zx}{1-zt}\right]  \,,\nonumber\\
& \left\langle A \right\rangle_6= \left\langle 1 \right\rangle =1,\nonumber %\\
\end{align}
$m_{\mu}$ is the muon mass $\left( P^{2}=-m_{\mu}^{2}\right) $ and
\begin{align}
& \quad t=\frac{(Q_{1}Q_{2})}{\left\vert Q_{1}\right\vert \left\vert Q_{2}\right\vert },
\quad x=\sqrt{1-t^{2}}\, ,\quad R_{i}=\sqrt
{1+\frac{4m_{\mu}^{2}}{Q_{i}^{2}}}\,,\nonumber\\
&\quad z=\frac{Q_{1}Q_{2}}{4m_{\mu}^{2}
}\left(  1-R_{1}\right)  \left(  1-R_{2}\right).  
\end{align}
$Z^{\mathrm{XY}}_{\mathbf{i},j}$ are polynomials in photon momenta
\begin{align}
&Z^{\mathrm{AA}}_{\mathbf{1},1}=(Q_1 Q_2) (Q_1^2+(Q_1  Q_2)) ,\quad
 Z^{\mathrm{AA}}_{\mathbf{1},2}=Q_2^2\frac{3Q_1^2+Q_2^2+Q_3^2}{4}         ,\quad\nonumber\\ 
&Z^{\mathrm{AA}}_{\mathbf{1},3}=-Q_1^2                       ,\quad
 Z^{\mathrm{AA}}_{\mathbf{1},4}=Q_2^2                        ,\quad
 Z^{\mathrm{AA}}_{\mathbf{1},6}=\frac{(Q_1  Q_2)-Q_2^2}{2} ,\quad\nonumber\\
&Z^{\mathrm{AA}}_{\mathbf{1},5}=Q_2^2 (2 m_{\mu}^{2}-Q_1^2-(Q_1  Q_2)) (Q_1^2+(Q_1  Q_2)), \quad
\end{align}
\begin{align}
&Z^{\mathrm{AA}}_{\mathbf{2},1}=\frac{Q_1^2(Q_1^2-2Q_3^2)}{2} ,\quad
 Z^{\mathrm{AA}}_{\mathbf{2},2}=\frac{Q_2^2(Q_2^2-2Q_3^2)}{2} ,\quad \nonumber \\
&Z^{\mathrm{AA}}_{\mathbf{2},3}=-Q_1^2                        ,\quad
 Z^{\mathrm{AA}}_{\mathbf{2},4}= Q_2^2                        ,\quad
 Z^{\mathrm{AA}}_{\mathbf{2},6}= Q_3^2-\frac{Q_1^2+Q_2^2}{2}                ,\quad\nonumber\\
&Z^{\mathrm{AA}}_{\mathbf{2},5}= Q_3^2 (Q_1^2 Q_2^2-2 m_{\mu}^{2} (Q_1  Q_2)) ,\quad   \nonumber
\end{align}

\begin{align}
&Z^{\mathrm{AB}}_{\mathbf{1},1}=-Q_1^2 Q_2^2((Q_1 Q_2)+Q_1^2) \frac{Q_3^2-Q_1^2}{2}                    ,\quad \nonumber\\
&Z^{\mathrm{AB}}_{\mathbf{1},2}=-Q_2^2\frac{(Q_1 Q_2)^2 (Q_2^2+(Q_1 Q_2))+Q_1^2 Q_2^2 Q_3^2}{2}   ,\quad \nonumber\\
&Z^{\mathrm{AB}}_{\mathbf{1},3}=-Q_1^2 (Q_1 Q_2) ((Q_1 Q_2)+Q_2^2)                               ,\quad \nonumber \\
&Z^{\mathrm{AB}}_{\mathbf{1},4}= Q_2^2 (Q_1 Q_2) ((Q_1 Q_2)+Q_2^2)                               ,\quad \\
& Z^{\mathrm{AB}}_{\mathbf{1},5}= Q_1^2 Q_2^4 Q_3^2\frac{(Q_1 Q_2)+Q_1^2}{2}                           ,\quad \nonumber\\
&Z^{\mathrm{AB}}_{\mathbf{1},6}= ((Q_1 Q_2)^2+Q_1^2 Q_2^2)\frac{(Q_1 Q_2)+Q_2^2}{2}                     ,  \nonumber
\end{align}
\begin{align}
&Z^{\mathrm{AB}}_{\mathbf{2},1}=\frac{Q_1^2}{2} \bigg(Q_2^2(Q_1^4 +(Q_1 Q_2) (Q_3^2-Q_1^2))+\nonumber\\&\quad +Q_1^2 ((Q_1 Q_2)((Q_1 Q_2)+5 Q_2^2)+2 Q_2^4)\bigg)   ,\quad\nonumber\\
&Z^{\mathrm{AB}}_{\mathbf{2},2}=\frac{Q_2^2}{2} \bigg(Q_1^2(Q_2^4 +(Q_1 Q_2) (Q_3^2-Q_2^2))+\nonumber\\&\quad +Q_2^2 ((Q_1 Q_2)((Q_1 Q_2)+5 Q_1^2)+2 Q_1^4)\bigg) ,\quad\nonumber\\
&Z^{\mathrm{AB}}_{\mathbf{2},3}= Q_1^2 ((Q_1 Q_2)+Q_1^2) ((Q_1 Q_2)+Q_2^2)                                                            ,\quad\\
&Z^{\mathrm{AB}}_{\mathbf{2},4}=-Q_2^2 ((Q_1 Q_2)+Q_1^2) ((Q_1 Q_2)+Q_2^2)                                                           ,\quad\nonumber\\
&Z^{\mathrm{AB}}_{\mathbf{2},5}=-Q_1^2 Q_2^2 Q_3^2 \frac{(Q_1 Q_2) (Q_2^2+Q_1^2)+2Q_1^2 Q_2^2}{2}                                     ,\quad\nonumber\\
&Z^{\mathrm{AB}}_{\mathbf{2},6}=\frac{(Q_1^2 Q_2^2-(Q_1 Q_2)^2) (Q_2^2+Q_1^2)}{2}-Q_1^2 Q_2^2 Q_3^2     .                   \nonumber
\end{align}

\end{document}